\documentclass[aps,pra,twocolumn,superscriptaddress,amsmath,amssymb,showpacs]{revtex4}
\usepackage{graphicx,color}
\usepackage{epstopdf}
\usepackage{epsfig}
\usepackage{soul} 
\usepackage{MnSymbol}

\newcommand{\beq}{\begin{equation}}
\newcommand{\eeq}{\end{equation}}
\newcommand{\bqa}{\begin{eqnarray}}
\newcommand{\eqa}{\end{eqnarray}}

\newcommand{\smallfrac}[2]{\mbox{$\frac{#1}{#2}$}}

\newcommand{\ket}[1]{\left|{#1}\right\rangle}

\newcommand{\half}{\smallfrac{1}{2}}

\definecolor{green}{rgb}{0.00,0.50,0.00}

\begin{document}

\title{Characterization of a qubit Hamiltonian using adaptive measurements in a fixed basis}

\author{Alexandr Sergeevich}
\affiliation{Centre for Engineered Quantum Systems, School of Physics, University of Sydney, Sydney 2006, Australia}
\author{Anushya Chandran}
\affiliation{ARC Centre for Quantum Computation and Communication Technology, and \\ Centre for Quantum Dynamics, Griffith University,
Brisbane, 4111, Australia} \affiliation{Department of Physics, Princeton University, Princeton, NJ 08544, USA}
\author{Joshua Combes}
\affiliation{ARC Centre for Quantum Computation and Communication Technology, and \\ Centre for Quantum Dynamics, Griffith University,
Brisbane, 4111, Australia}
\affiliation{Center for Quantum Information and Control, University of New Mexico, Albuquerque, NM 87131-0001, USA}
\author{Stephen D. Bartlett}
\affiliation{Centre for Engineered Quantum Systems, School of Physics, University of Sydney, Sydney 2006, Australia}
\author{Howard M. Wiseman}\email{H.Wiseman@griffith.edu.au}
\affiliation{ARC Centre for Quantum Computation and Communication Technology, and \\ Centre for Quantum Dynamics, Griffith University, Brisbane, 4111, Australia}

\date{15 November 2011}

\begin{abstract}
We investigate schemes for Hamiltonian parameter estimation of a
two-level system using repeated measurements in a fixed basis.  The simplest (Fourier based) schemes yield an estimate with a mean square error (MSE)  that decreases at best as a power law $\sim N^{-2}$ in the number of measurements $N$. By contrast, we present numerical simulations indicating that an adaptive Bayesian algorithm, where the time between measurements can be adjusted based on prior measurement results, yields a MSE which appears to scale close to $\exp(-0.3 N)$. That is, measurements in a single fixed basis are sufficient to achieve exponential scaling in $N$.
\end{abstract}


\maketitle

\section{Introduction}

Efficient methods for the characterization of quantum systems to
extremely high precision are important both to reach new regimes of
physics and to build robust quantum technologies~\cite{mikeike,WisMilBook}.  One of the most
fundamental characterisation tasks is the estimation of the
parameters of a Hamiltonian in a two-level system.  Several previous
studies~\cite{cole, schirmer,RalComWis10} used a method of repeatedly
initializing the two-level system and then performing measurements
in a fixed basis after consecutively longer intervals (during which
the system evolves under its Hamiltonian) and then averaging many runs. By calculating the
Fourier transform of the resulting signal and identifying its peak,
it is possible to obtain an estimate for the rate of evolution, and thus the desired Hamiltonian parameter.

This approach is noticeably more efficient (faster) in practice than
quantum process tomography~\cite{Tom1}, requiring only
measurements in one particular basis (as state initialization can be
done via measurement).  However, it still demands a large number of
measurements for moderate accuracy. For example, in
Ref.~\cite{cole}, the two parameter estimation procedure required at
least $10^{6}$ measurements in order to reach a joint variance of
$10^{-3}$ in the parameters being estimated. Such large numbers of
measurements can pose a problem, especially in solid-state systems
where the measurement time is typically the slowest timescale,
often many orders of magnitude longer than the period for coherent
evolution. To specifically address such situations, we quantify
resources in our estimation schemes as $N$, the number of measurements 
used, rather than the total evolution time as is commonly used in
phase estimation schemes using optics and assuming instantaneous
measurements~\cite{WisemanNature} (however cf.~\cite{Giedke}). 
We note however that our techniques could easily be modified to take into account both the waiting time and the measurement time.

We emphasise that, unlike schemes based on the quantum phase estimation algorithm~\cite{QPEA,mikeike} such as that proposed in Refs.~\cite{Giedke,BoixoSomma2008}, we restrict our measurement to a fixed basis and do not allow any controlling unitary dynamics. That is, our schemes are limited to preparing a pure state in this fixed basis, evolving for some time under the Hamiltonian, and measuring in this same basis. The motivation for this restriction is simple:  in most situations, the unitary required to change bases would be generated by the very Hamiltonian parameter that we are attempting to estimate.

A motivating example is provided by recent experimental progress in
the development of spin qubits in semiconductor quantum dots,
specifically, GaAs double dot systems where a qubit is
defined using two electron spins in a singlet/triplet
configuration~\cite{Hanson}.  With one electron in each dot, the
states $|\!\!\uparrow\downarrow\rangle$ and $|\!\!\downarrow\uparrow\rangle$ experience an energy splitting
proportional to the difference in the $z$-component of the
magnetic field, $\Delta B^z$, resulting from the hyperfine
interaction with nearby lattice nuclear spins.  Because variations
in $\Delta B^z$ are the primary source of decoherence in these spin
qubits, there has been considerable recent interest in the
measurement and control of this nuclear magnetic field by using the
spin qubit as both a probe and feedback
mechanism~\cite{Coish,Reilly,Foletti,Bluhm}.  In addition, a well-known
and stable value of this field can serve as a source of coherent
quantum operations (i.e., logic gates) on the spin
qubit~\cite{Foletti,Bluhm}. (However, one cannot use this effect to
change the measurement basis and implement a quantum phase
estimation algorithm as in~\cite{QPEA,mikeike,Giedke} without first
estimating the field; thus, our requirement for fixed basis
measurements.) With the recent demonstration of single-shot
projective measurements of the spin qubit~\cite{Barthel}, parameter
estimation of $\Delta B^z$ in such systems is possible~\cite{Bluhm}.  The system
coherently evolves on a nanosecond timescale, whereas the
measurement time is $\sim\!\!10\mu$s~\cite{Foletti,Johnson,Stopa}.  (In these systems the coherent evolution is switched off during the measurement process.)  For this estimation problem, then, we seek schemes that minimize the number of measurements required for a given accuracy.

In this paper, we consider the performance of a range of schemes for
such a parameter estimation, using numerical simulations. First, we demonstrate that a Bayesian approach outperforms the Fourier estimation techniques.  We show that, while schemes using a predetermined sequence of measurements yield a mean square error (MSE) decreasing polynomially in the number of measurements, a drastic
improvement can be found by using \emph{adaptive}
measurement approach~\cite{WisMilBook}.  The adaption is done by \emph{local optimization} \cite{BerryWiseman2000}: 
the time intervals between preparation and measurement is chosen  to minimize the
expected MSE, conditioned on the result of that future measurement.  Numerical simulations for the adaptive scheme are consistent with exponential scaling in $N$ for the MSE 
in the estimate of the parameter, while the best 
nonadaptive algorithm found has a power law scaling in $N$. Our result
demonstrates that exponential scaling of the estimate precision  
can be achieved with a single, fixed basis of
measurement, rather than requiring measurements in arbitrary bases
as in the quantum phase estimation
algorithm~\cite{QPEA,mikeike,Giedke}. Finally, we show that
quite good performance is achievable by a
\emph{locally optimal non-adaptive} scheme.  



\section{The Problem}
\label{par_est}
We consider the problem of estimating a single unknown
parameter of a qubit Hamiltonian, of the form $H= \omega
\sigma_{z}/2$. To simplify later calculations we assume that
$\omega$ is a random parameter uniformly distributed over the interval
$[0,\omega_{0}]$, where $\omega_{0}$ is the largest
possible value of $\omega$. In order to estimate $\omega$, we probe
that system with projective measurements of the $x$ component of
spin at different times.  (Note that this measurement basis is
\emph{not} the energy eigenbasis of the Hamiltonian; otherwise,
parameter estimation would not be possible.) We initialize the state
as an eigenstate $\ket{+}$ of $\sigma_{x}$ at $t=0$; we note that this initialization is naturally performed at each step by the previous measurement. The Nyquist--Shannon theorem 
suggests that we want to choose the time between preparation and
measurement to be as small as $\tau \equiv \pi/\omega_0$.  This minimum time interval $\tau$ (and hence maximum parameter range $[0,\omega_0]$) will be determined by experimental considerations, and it is therefore reasonable 
to assume that the {\em waiting time} between the $k$th preparation and the $k$th measurement, 
is an integer multiple of $\tau$. That is, $t_k = m_k \tau$. 
The Hamiltonian in this case generates the time evolution
\begin{equation}
 \ket{\psi(t)} = \cos\left( \omega t/2 \right)\ket{+} - i\sin\left( \omega t /2\right)\ket{-}\,,
 \label{red}
\end{equation}
and the probabilities for the outcomes of the $k$th measurement are
\begin{equation}
p_k({+}|\omega) = \cos^2 \left(\frac{\pi \omega
m_k}{2\omega_0} \right),  \quad p_k({-}|\omega) = \sin^2
\left(\frac{\pi \omega m_k}{2\omega_0} \right). \label{prob}
\end{equation}
The relevant resource in our estimation procedure is the number of
measurements. The problem then becomes:  given a fixed number of
measurements $N$, how should one proceed in determining the waiting times 
$m_k\tau$, and how does one infer $\omega$ from
the results of the measurements?

This problem falls within the domain of \emph{quantum parameter estimation}, wherein one seeks to identify 
an unknown parameter influencing the preparation or dynamics of a quantum system. 
 The  canonical example is estimating the phase of a unitary 
operator, which is closely related to characterizing 
a Hamiltonian with an unknown magnitude, as in this paper. Quantum parameter estimation techniques can allow for high-precision phase estimation below the classical (shot noise limit) as well as power algorithms
for quantum computation~\cite{QPEA,mikeike}.  In quantum parameter estimation problems,
such as the one we consider, it is necessary to carefully tailor measurements
and process their outcomes in order to make inferences on the (unaccessible)
parameter of interest.  While many techniques for quantum parameter estimation make use of entanglement~\cite{GLM}, it is in some situations possible to replace entangled states with repeated application of the unknown unitary on a single system 
prior to measurement~\cite{WisemanNature,Higgins2009,Berry2009}. Adaptive measurements--- 
which have been proposed and used for quantum parameter estimation~\cite{Wis95c,BerryWiseman2000,Arm02,WisemanNature,Giedke,BoixoSomma2008,Xiang2011}, 
quantum tracking~\cite{Whe10,KarWis11}, 
 state discrimination~\cite{JM07,Higgins09}, state estimation~\cite{Fischer2000,Hannemann2002,Check2007}, and 
 quantum computing~\cite{KLM01,RauBri01,Prevedel2007}---can play a key role in this context 
because of the phase ambiguity inherent in estimating a parameter that appears in the problem only as the 
scale of an anti-Hermitian operator 
(i.e. $-iHt$) which is exponentiated~\cite{WisemanNature,Berry2009}.

One question of interest in quantum parameter estimation is how close the measurement comes to the 
so-called Heisenberg limit \cite{WisMilBook}. This is the limit on the variance, or Fisher information, 
of the unknown parameter, imposed by Heisenberg's uncertainty principle. 
For example, in the case of phase estimation, this limit scales 
as $N_U^{-2}$, where  $N_{U}$ is the total number of times  
the unitary is applied, whether it is applied $N_{U}$ times to a single system, or once each across $N_{U}$ systems, 
or anything in between \cite{GLM,Berry2009}. Restricting to a single system (as in our paper), in the asymptotic limit 
$N_U\to\infty$, the run-time of any experiment that can attain the Heisenberg limit will scale as $N_U$. 
 However, as noted above, in the practical regime of interest to us, the run-time of the experiment will be determined by 
$N$, the number of preparation and measurement steps, not the evolution time $\sim N_U$ between preparation and measurement.
Since there is no fixed relation between $N_U$ and $N$ (except that $N_U \geq N$), 
the Heisenberg limit does not automatically translate into any limit on the variance as a function of $N$. 
Rather, we must determine how well various schemes scale with $N$ (in the regime of interest), 
and thereby determine the best of them.


\section{Schemes for parameter estimation}

In the following sections, we present techniques for Hamiltonian parameter estimation based on Fourier methods (A), 
 and then those based on Bayesian methods (B--D).  The latter include simple non-adaptive Bayesian schemes (B), a locally optimal adaptive Bayesian scheme (C), and a locally-optimal non-adaptive Bayesian scheme (D).

\subsection{Fourier Estimation Techniques}
\label{fourier_est}
A simple strategy for this problem is to measure at uniformly distributed times $t_k = k \tau$, i.e., to choose $m_k = k$. The set of measurement results constitutes a measurement record, and can be loosely thought of as
one realization of a random process \cite{RalComWis10}.  One method to estimate the parameter is to Fourier transform the measurement record and identify the peak of the spectrum as the best estimate for $\omega$~\cite{cole, schirmer}.  However, this is not the only strategy.  For example, for each $t=m\tau$ we could prepare, evolve and measure twice, with the range $m\in \{1,\ldots, N/2\}$. The resulting measurement record can be viewed as two realizations of the random process, and averaging these two realizations will reduce the effect of projection noise (i.e., the noise due to the indeterminacy of the measurement outcomes).  We can define a family of schemes, wherein $M$ different choices of waiting times are each repeated $n$ times, with a total of $N=nM$ measurements.

\begin{figure}
\begin{center}
\includegraphics[width=1\hsize]{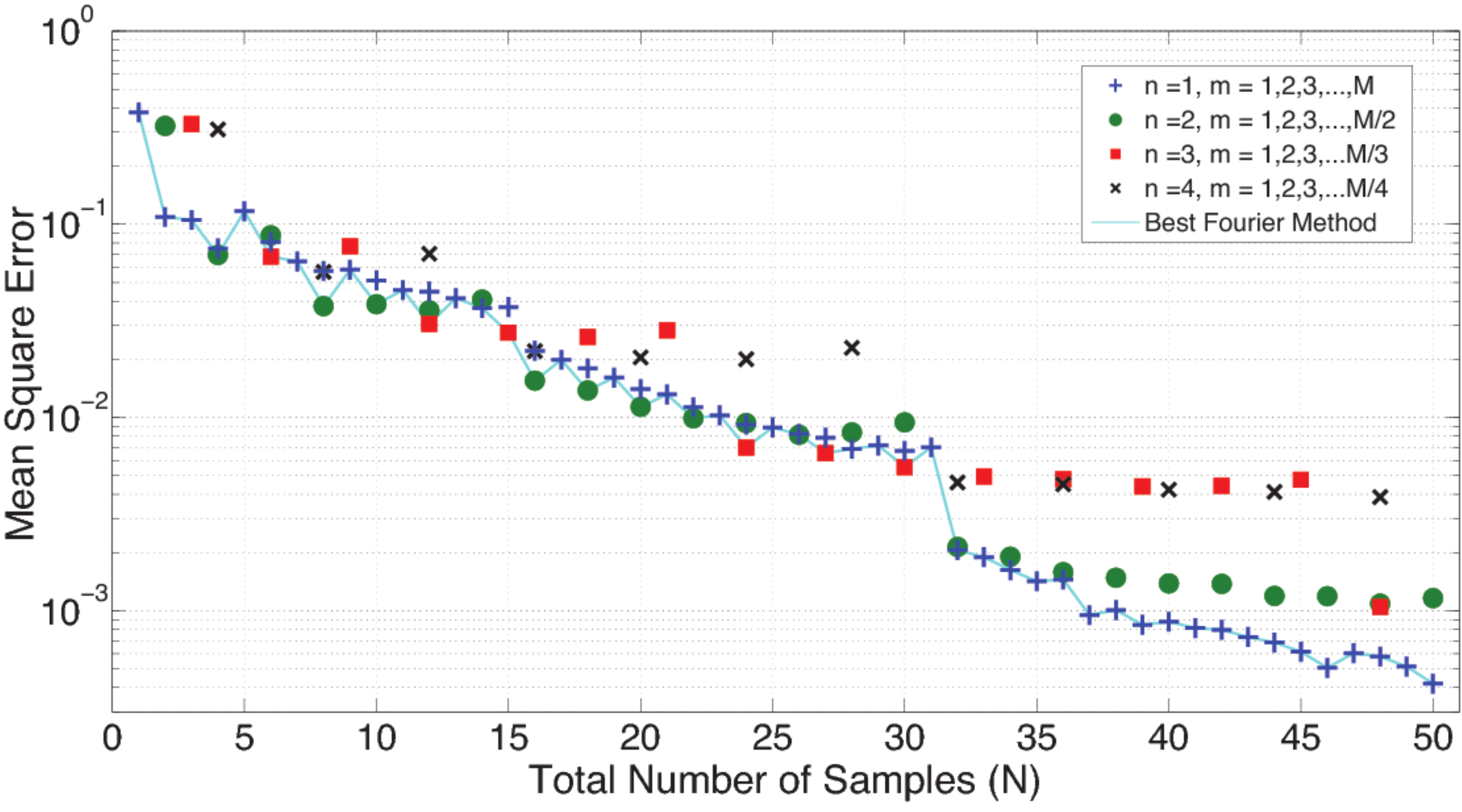}
\caption{The MSE in the frequency
estimate for Fourier-based schemes for some different partitionings of $N$.  The ``Best Fourier
Method'' (solid line) for a particular $N$ is given by the best performing scheme for that $N$. To generate the data for this plot each scheme was sampled $100\,000$ times, consequently the error bars (not shown)
are slightly smaller than the markers for the data points.}\label{standard}
\end{center}
\end{figure}

Using this technique we have considered partitions of $N$ where
$n\in\{1,2,3,\ldots,10\}$.  We find that $n\in \{1,2,3\}$ give the best
MSE scaling depending on the value of $N$; see Fig.~\ref{standard}. In what follows, we use the partitioning that minimizes the MSE for a given $N$, and call this method the \emph{best
Fourier method}.  The MSE in the estimate of
$\omega$ as a function of $N$ for this method sets the
benchmark to which our  more sophisticated schemes will be compared. 
For a large number of measurements $N$, the scaling of the MSE is found to have power-law scaling in $N$ with a power close to $-2$.  Specifically, for a fit of $N$ from $36$ to $3000$ with each point sampled $400$ times, the
$95\%$ confidence interval in the power is $(2.096, 2.064)$, $R^2 = 0.956$. 

\subsection{Bayesian Parameter Estimation}
\label{bayes_est}
We now consider performing a Bayesian analysis of the same schemes
described above.   Here, one's knowledge about the unknown parameter
$\omega$ is represented as a probability distribution
$P(\omega)$. Using the mean of this distribution as one's 
estimate gives a MSE which is equal to the variance of this probability distribution,
averaged over all realizations. Thus we can use the variance of the posterior (i.e. post-measurement) 
$P(\omega)$ as the measure 
of precision for Bayesian schemes.  

We take the uniform prior probability distribution $P_{0}(\omega)= 1/\omega_0$.  At each measurement, the outcomes $+$ and $-$ are expected with probabilities given by
Eq.~(\ref{prob}) and so the total probability distribution can be
updated each measurement step using Bayes rule. The general
expression for conditioned probability distribution of $\omega$ given $k$ measurements, is 
\begin{equation}
  \label{product} P_k (\omega|r_k\ldots r_1) =
\mathcal{N} \prod_{j=1} ^k \bigl[ 1  + r_j \cos(\pi m_j \omega /
\omega_0)\bigr],
\end{equation}
where $r_{j}=1$ ($r_j = -1$) corresponds to learning that the $j$th
result is $+$ ($-$), $m_{j}$ denotes the
waiting time for the $k$th measurement, and $\mathcal{N}$ is a
normalization constant.  These conditional probabilities allow for
any Bayesian parameter estimation task to be performed on a given
measurement record. 

We can avoid the computationally costly practice of discretizing the distribution in $\omega$ by using the following technique.  As Eq.~(\ref{product}) is an even function of $\omega$, it can be represented as a Fourier cosine series:
\begin{equation}
  \label{sumo}
  P_{k} (\omega|r_k \ldots r_1) = \half c_{k}(0)+\sum_{q=1} ^K c_{k}(q) \cos (q \pi\omega/\omega_0),
\end{equation}
where $K=\sum_{j=1}^{k} m_j$. The distribution is normalized by dividing
by $\half \omega_0 c_{k}(0)$.  Because the number of terms in Eq.~(\ref{sumo}) is finite, the representation of the distribution is exact.
It is then possible to derive an analytic expression for $\langle\omega\rangle= \int _{0}^{\omega_0} \omega P_{k}(\omega|r_{k}\ldots r_{1})d\omega$ 
(and $\langle\omega^{2}\rangle$) using the Parseval theorem, which relates the convolution of two functions, $ P_{k}(\omega|r_{k}\ldots r_{1})$ and $\omega$ (or $\omega^{2}$), to the summation of the product of their corresponding Fourier coefficients. Performing this calculation gives an explicit formula for the variance, $V = \langle\omega^2\rangle-\langle\omega\rangle^2$, given by
\begin{multline}
  \label{variance}
  V = \left [ \frac{\omega_{0}^2}{3}+\sum_{q=1}^K\frac{2c_q\omega_{0}^2(-1)^q}{c_0(q  \pi)^2}\right ] \\
  -\left [\frac{\omega_{0}}{2}+\sum_{q=1}^K\frac{c_q\omega_{0}[(-1)^q-1]}{c_0(q\pi)^2}\right ]^2.
\end{multline}

\begin{figure}
\begin{center}
\includegraphics[width=1\hsize]{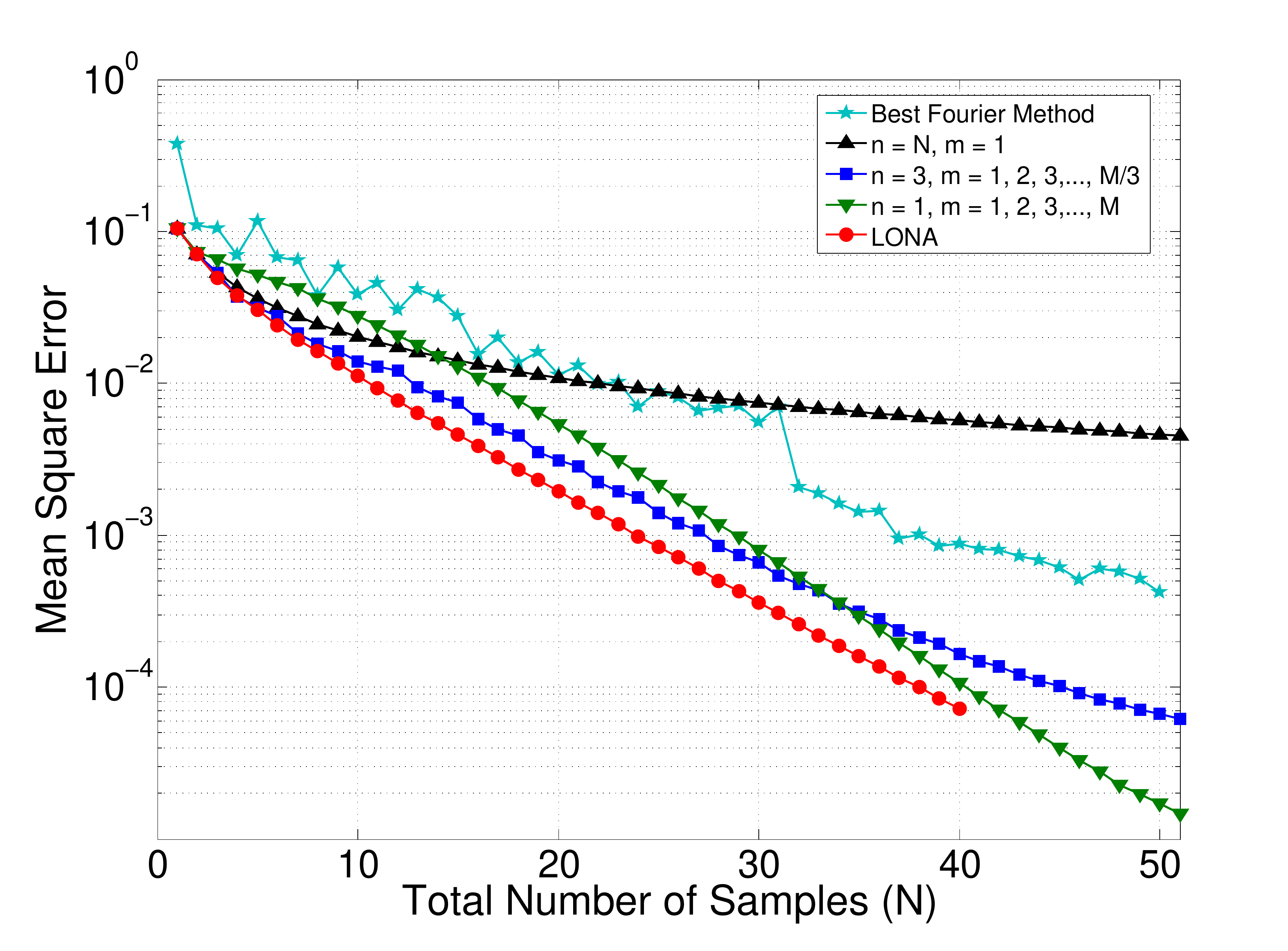}
\caption{A comparison of the scaling of the variance for Fourier and
non-adaptive Bayesian methods. The stars are for the Best Fourier method
described in the text. The upward triangles are for the Bayesian
method where the $N$ samples are taken at at $m=1$. The downward
triangles are for the Bayesian $n=1$ method where the $N$ samples are
sequentially taken at times $m=1,2,3...M$ (here $M=N$). The squares are for the
Bayesian $n=3$ method where each point is sampled three times, i.e., 
$m=1,1,1,2,2,2,\ldots M/3$.  The circles are for the locally optimal non-adaptive (LONA) method.} \label{nonadplot}
\end{center}
\end{figure}

With this technique, we can perform a Bayesian analysis of the measurement outcomes of the predefined parameter estimation schemes discussed above. We simulated over $100\,000$ runs for $n\in\{1, 2, 3\}$ and $N$ up to $36$ measurements to obtain the variance as a function of $N$ shown on Fig.~\ref{nonadplot}.  
The simplest algorithm with $n=N$, $m=1$ yields an asymptotic scaling of the variance  $\sim N^{-1}$. This is obtained by fitting the first $5000$ steps ($95\%$ confidence interval of power law scaling $(0.9893, 0.9895)$, with $R^2 > 0.9999$).  By contrast, algorithms with $n\in\{1, 2, 3\}$ that uniformly distribute measurement times exhibit an improved error scaling of the variance $\sim N^{-3}$ (for $n=1$, the first $1000$ steps yield the $95\%$ confidence interval for the power to be $(3.0376, 3.0384)$, with $R^2 > 0.9999$). This latter result demonstrated a significant improvement over the Fourier analysis using the same sequences, an improvement due solely to the superior data processing of Bayesian analysis. We find that $n=3$ is the most effective for smaller $N$ (likely due to a reduction in projection noise), whereas $n=1$ has a better scaling for large $N$.

\subsection{Adaptive Bayesian Scheme}
\label{adapt_est}
We next consider an adaptive method in which the waiting times $t_k$ are chosen based on previous results.  Specifically, we adaptively chose $m_k$ so that the conditional expectation of the variance $\mathrm{E}[V]$ after the measurement is minimized.
The conditional expectation of the variance after the $k$th result
for the waiting time $m_k$ is
\begin{equation}
  \label{vbar}
  \textrm{E}[V_{k}|m_k]=\frac{  {c_{k}(0|{+})} V_{k|{+}}  + {c_{k}(0|{-})} V_{k|{-}}   }{c_{k}(0|{+})+c_{k}(0|{-})}
\end{equation}
where $c_{k}(0|r)$ denotes the Fourier coefficient $c_{k}(0)$ of
Eq.~(\ref{sumo}) given the measurement outcome $r$ from the $k$th measurement
and $V_{k | r}$ is the variance conditional on this outcome.  
The factors multiplying each  $V_{r}$ are the
probabilities of the outcomes. Our strategy is to, at each step,
choose the waiting time $\tau m_k$   in order to minimize
the expected variance of posterior distribution
$\textrm{E}[V_{k}|m_k]$.

\begin{figure}
\includegraphics[width=1\hsize]{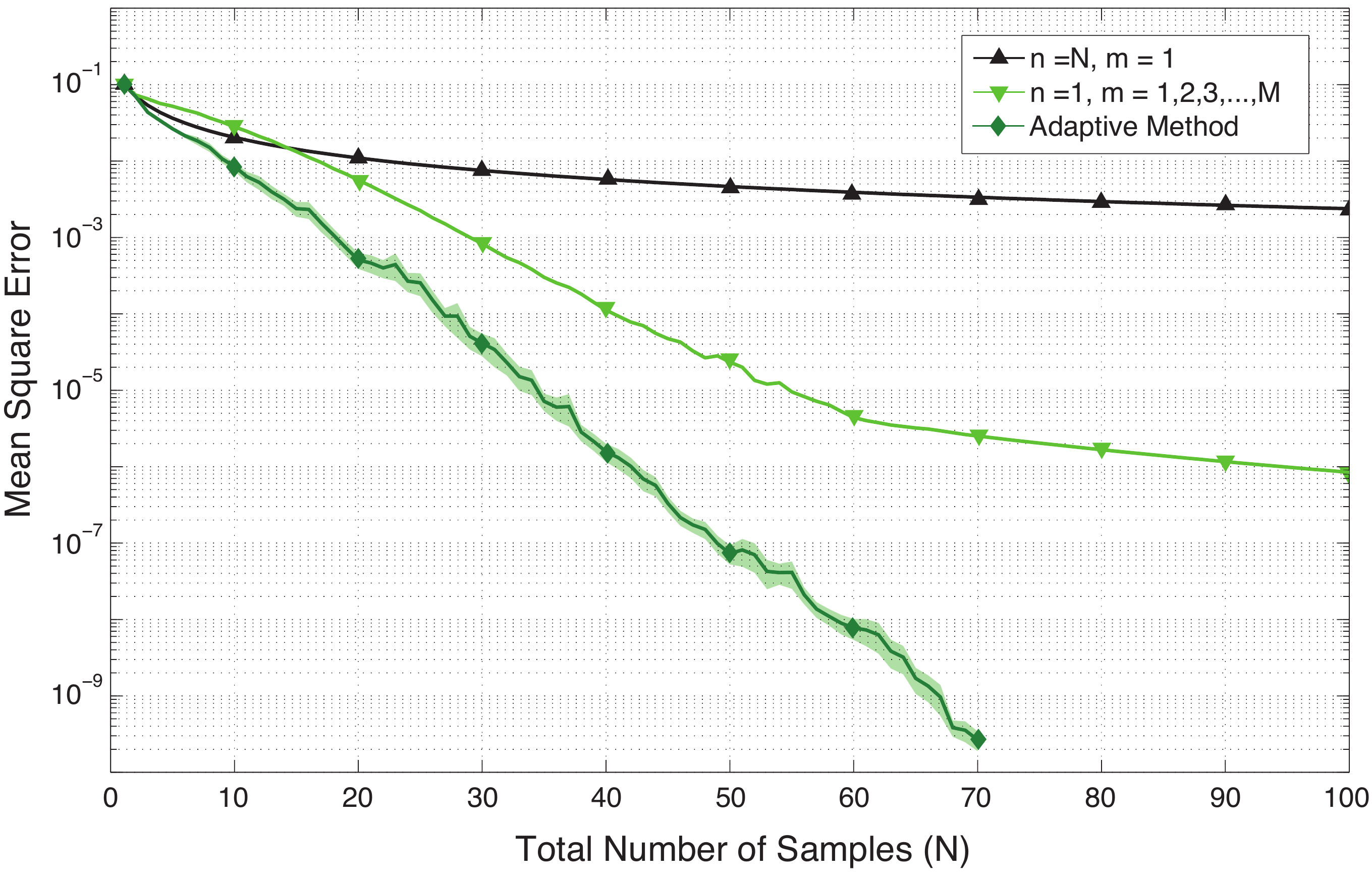}
\caption{The plots for mean square error as a function of the total number of measurements:  ($\filledmedtriangleup$) the Bayesian method where the $N$ samples are taken at at $m=1$; ($\filledmedtriangledown$) the Bayesian method where the $N$ samples are sequentially taken at times $m=1,2,3,\ldots,M$; ($\filleddiamond$) the locally optimal adaptive strategy.  The average MSE was computed from $10\,000$ simulations and the largest possible allowed waiting time was $m_{max}=1000$.  The shaded region on the adaptive strategy plot represents the standard deviation of the MSE.} \label{adplot}
\end{figure}

Fig.~\ref{adplot} shows the scaling of the variance as a function of
$N$ for this adaptive strategy, compared with the above non-adaptive
Bayesian schemes.  Unlike the previous schemes where we could fit numerics beyond $N=200$, here we are limited up to $N=70$.  Fitting to the first $70$ measurements, the MSE of this adaptive strategy scales
as $\sim \exp(-a N)$ with $a=0.2990$ (with $95\%$ confidence interval for exponent $a$ of $(0.2796, 0.3185)$) and $R^2=0.9847$. We compare this exponential fit with the best power fit, which gives $\sim N^{-7}$ with $R^2$ of only $0.9064$.   That is, we have obtained an exponentially decreasing MSE similar to that of Ref.~\cite{Giedke} without the need to alter measurement bases throughout the protocol. 

While scaling in the number of steps can be exponential in $N$, this does not mean breaking the Heisenberg limit on the variance, scaling as $N_U^{-2}$. The reason is that the adaptive algorithm (as well as other previously discussed schemes) require exponentially longer waiting times for large number of steps, so that $N_U$ varies exponentially with $N$. As noted above, in the truly asymptotic regime, the evolution time will become much longer than the measurement time. In that limit 
our algorithm will scale worse than the Heisenberg limit in terms of total run-time, since it is optimized for a different problem.



\subsection{Locally Optimal Non-Adaptive Scheme}\label{lona_est}
Due to the computation required in optimally choosing the waiting time at each measurement step, the 
complexity of the adaptive scheme could present problems for practical use.  We therefore seek to identify non-adaptive schemes with the best possible performance. Several heuristics enable one to design such schemes; we will describe one.

In the initial step, we begin with a flat prior and determine what
waiting time will minimize $E[V_{1}|m_{1}]$.  For the second
measurement, we determine the optimal waiting time to minimize
$E[V_{2}|m_{2}]$ given that a measurement was performed at $m_{1}$
but the result is not known.  This process is then repeated. We can
find the first 20 steps analytically --- i.e., $\{m_{1},m_{2},m_{3},m_{4},m_{5},\cdots\}=\{1,1,2,1,3,\cdots\}$.
After that, we use a numerical search. Because these waiting
times are determined from expected rather than actual statistics, it
is non-adaptive; this string of waiting times is determined offline.
We denote this scheme the locally optimal non-adaptive (LONA)
scheme. Because of computational complexity, it becomes intractable
to determine the error scaling of LONA for large number of steps. 
However this algorithm performs well for small $N$ (see 
Fig.~2), and so is appealing in situations where relatively few measurements are 
necessary and adaptive methods are not feasible.





\begin{table}
\begin{tabular}{|l|c|c|}
\hline Algorithm & Steps to $V = 10^{-3}$ & Steps to $V = 10^{-5}$ \\
\hline 
Bayesian $n=N$ & $242$ & $\gtrsim 2\cdot 10^{4}$  \\
Fourier & $33$ & $\gtrsim 130$ \\
Bayesian $n=1$ & $29$ & $55$  \\
LONA & $24$ & $~49 $  \\
Adaptive & $20$ & $35$  \\ \hline 
\end{tabular}
\caption{Comparison of schemes.  Number of measurements required to meet a desired variance of $10^{-3}$ and $10^{-5}$.}\label{tab:Table}
\end{table}

\section{Discussion}
We have shown that Bayesian methods can be used for efficient
Hamiltonian parameter estimation schemes.  Our adaptive Bayesian
algorithm, which is locally optimal, provides an exponential 
improvement in the scaling of the variance with the number of measurements
performed, and unlike methods based on the quantum phase estimation
algorithm does not require adaptive measurement bases --- the 
measurements are in a fixed basis and only the waiting times between 
them are adapted.  See Table~\ref{tab:Table} for a comparison of schemes.  

We note that decoherence will in general affect the performance of these schemes. 
Recently, considerable progress has been made in the understanding of 
how to determine asymptotic limits in parameter estimation in the presence of decoherence~\cite{Escher11}.  
 While a detailed analysis of the effects of decoherence is beyond the scope of this work, we note that simulations based on realistic parameters for GaAs double dot spin qubits possessing coherent evolution on the nanosecond timescale and dephasing times of microseconds demonstrate only a small effect on the performance of the LONA scheme up to $\sim 30$ measurements. 

We emphasise that, in our analysis, we have used the number of measurements $N$ to represent the resource cost of the scheme; this differs from typical phase estimation scenarios, where $N$ represents the total number of applications or probes (e.g., number of photons) of the Hamiltonian~\cite{GLM}.  As such, the scalings of our various schemes cannot be directly compared with other results, nor the terminology based around the standard quantum limit or the Heisenberg limit.  For a simple comparison, it should be noted that the waiting time in our schemes typically becomes exponetially long for large $N$, and so even the adaptive Bayesian scheme with its exponential scaling in terms of number of measurements $N$ will appear Heisenberg-limited when total time is used instead.

\emph{Note added in proof:}  Recently a preprint~\cite{Ferrie11} has appeared which re-examined many of our schemes using analytical techniques, confirming the scaling laws we obtained numerically.

\begin{acknowledgments}
The authors would like to acknowledge fruitful conversations with Jared H. Cole, Jason F. Ralph, and David Reilly.  This research was supported by the Australian Research Council Centre of Excellence scheme (project numbers CE110001027 and CE110001013).  SDB acknowledges support from ARO/IARPA project W911NF-10-1-0330. JC acknowledges support from National Science Foundation Grant No. PHY-0903953 and Office of Naval Research Grant No. N00014-11-1-008.
\end{acknowledgments}

\bibliographystyle{h-physrev3}
\end{document}